\documentclass[prc,aps,showpacs,floatfix,nofootinbib,twocolumn,letterpaper]{revtex4}
\usepackage{graphicx}
\usepackage{amssymb}
\usepackage{amsmath}
\usepackage{amsfonts}
\usepackage{array}
\usepackage{dcolumn}
\usepackage{comment}

\makeatletter

\newcommand{\Rmnum}[1]{\expandafter\@slowromancap\romannumeral #1@}

\newcommand{\lb}{\langle}
\newcommand{\rb}{\rangle}
\newcommand{\be}{\begin{equation}}
\newcommand{\ee}{\end{equation}}
\newcommand{\ba}{\begin{eqnarray}}
\newcommand{\ea}{\end{eqnarray}}
\newcommand{\ad}{a^\dagger}
\newcommand{\Tr}{{\rm Tr}}

\newcommand{\bi}{\bibitem}
\newcommand{\ma}{\mathcal{A}}
\makeatother

\begin{document}
\title{Level densities of nickel isotopes: microscopic theory versus experiment}
\author{M. Bonett-Matiz$^1$, Abhishek Mukherjee$^{2,1}$, and Y. Alhassid$^1$}
\affiliation{$^1$Center for Theoretical Physics, Sloane Physics
 Laboratory, Yale University, New Haven, CT 06520\\
$^2$ECT*, Villa Tambosi, I-38123 Villazzano, Trento, Italy}
\begin{abstract}
We apply a spin-projection method to calculate microscopically the level densities of a family of nickel isotopes $^{59-64}$Ni using the shell model Monte Carlo approach in the complete $pfg_{9/2}$ shell. Accurate ground-state energies of the odd-mass nickel isotopes, required for the determination of excitation energies, are determined using the Green's function method recently introduced to circumvent the odd particle-number sign problem.  Our results are in excellent agreement with recent measurements based on proton evaporation spectra and with level counting data at low excitation energies. We also compare our results with neutron resonance data, assuming equilibration of parity and a spin-cutoff model for the spin distribution at the neutron binding energy, and find good agreement with the exception of $^{63}$Ni.
\end{abstract}
\maketitle

{\em Introduction.}
Nuclear level densities play an important role in determining the elemental abundances in stellar nucleosynthesis~\cite{Clayton1968, Rauscher1997}. However, the microscopic calculation of level densities from underlying effective interactions has been a major theoretical challenge~\cite{Grawe2007}.

The configuration-interaction (CI) shell model approach~\cite{Caurier2005} offers an attractive framework for reliable calculation of nuclear level densities.
 In this approach, both shell effects and correlations are included \emph{a priori}; thereby eliminating the need for \emph{a posteriori} parameter
fitting as is done, for example, in empirical approaches based on the back-shifted Bethe formula (BBF)~\cite{Bethe1936, Gilbert1965, Ripl3}.

However, the conventional shell model approach, which is based on direct diagonalization of the Hamiltonian in a truncated CI shell model space,
 becomes prohibitively difficult in medium-mass and heavy nuclei because of the combinatorial increase of the dimensionality of the many-body Hilbert space.
An alternative is provided by the shell model Monte Carlo (SMMC) approach~\cite{Lang1993,Ormand1994, Alhassid1994,Koonin1997,Alhassid2001}. The SMMC can be used to calculate thermodynamic observables within the CI shell model framework. In particular, it has been shown to provide accurate estimates of nuclear  state densities~\cite{Nakada1997, Nakada1998, Ormand1997,Langanke1998, Alhassid1999, Alhassid2007,Ozen2007}. The computational resources required for SMMC calculations scale gently with the size of the single-particle space, enabling calculations in very large many-particle model spaces.

A variety of methods are used to extract level densities from experimental data: direct level counting at very low excitation energies, neutron or proton resonance data at the neutron or proton binding energy, and Ericson's fluctuation analysis~\cite{Richter1974} at higher excitation energies ($E_x \gtrsim 15$ MeV in medium-mass nuclei).  Level densities at intermediate energies can be extracted from charged particle reactions~\cite{Lu1972}, and using the Oslo method~\cite{Oslo2000} in nuclei for which both level counting and neutron resonance data are available.  More recently, neutron and proton evaporation spectra have been used to determine level densities at intermediate energies independently of the neutron resonance data~\cite{Voinov2009}.
In particular, the level densities of a family of nickel isotopes ($^{59-64}$Ni)  were extracted from proton evaporation spectra~\cite{Voinov2012}.

The usual density calculated in the SMMC method is the {\em state} density, in which the spin degeneracy factor $2J+1$ of each level with spin $J$ is taken into account. However, experiments in the intermediate excitation energy regime often measure the {\em level} density, in which each level is counted only once irrespective of its spin degeneracy.  In this work we calculate the level densities of nickel isotopes ($^{59-64}$Ni) directly with the SMMC method using the spin-projection technique described in the recent work of Ref.~\onlinecite{Alhassid2013}. This method was applied to even-mass nuclei only and here we present its first application to odd-mass nuclei. The ground-state energies of the odd-mass nickel isotopes are calculated accurately using a method recently introduced in Ref.~\onlinecite{Mukherjee2012}.

We find that our theoretical level densities agree very well with the experimental level densities extracted recently from proton evaporation spectra~\cite{Voinov2012} and with level counting at low excitation energies. We also compare our SMMC results with level densities extracted from neutron resonance data at the neutron binding energy, assuming equal densities of positive and negative parity levels and a spin cutoff model with rigid-body moment of inertia. We find good agreement except for $^{63}$Ni.

\emph{The shell model Monte Carlo approach.}
The SMMC method is based on a representation of the Gibbs operator $e^{-\beta\hat H}$ (where $H$ is the CI shell model Hamiltonian and $\beta=1/T$ is the inverse temperature)  as a functional integral over one-body propagators $\hat U_\sigma$ describing nucleons moving in external time-dependent auxiliary fields $\sigma=\sigma(\tau)$. This is formally expressed in terms of the Hubbard-Stratonovich (HS) transformation~\cite{Hubbard1959}
\be\label{eq:HSTrans}
e^{-\beta\hat H}=\int D[\sigma]G_\sigma\hat U_\sigma
\ee
where $G_\sigma$ is a Gaussian factor.

The thermal average of an observable $\hat O$ can be written as
\be
\label{eq:expVal}
\lb \hat O \rb = \frac{\int D[\sigma]W_\sigma\Phi_\sigma
 O_\sigma}{\int D[\sigma]W_\sigma\Phi_\sigma}\;,
\ee
where $W_\sigma = G_\sigma|\mathrm{Tr}\text{ }\hat U_\sigma|$ is a positive definite weight, $\Phi_\sigma = \mathrm{Tr}\text{ }\hat U_\sigma/|\mathrm{Tr}\text{ }\hat U_\sigma|$ is the Monte Carlo sign function and  $O_\sigma=\Tr \,(\hat O \hat U_\sigma)/\Tr\,\hat U_\sigma$. In SMMC, the r.h.s.~of Eq.~(\ref{eq:expVal}) is calculated using stochastic sampling of the auxiliary-field  configurations $\sigma$ from the distribution $W_{\sigma}$. Thermal averages at fixed number of protons and neutrons are calculated in the canonical ensemble with the help of particle-number projection.

The thermal energy $E=E(\beta)$ is calculated from $E(\beta)=\lb \hat H \rb$ and the partition function $Z (\beta)$ is obtained by integrating the thermodynamic relation $-{d\ln Z/ d\beta}= E$. The entropy and heat capacity are calculated from $S = \ln Z + \beta E $ and
$C=-\beta^2 d E /d \beta$, respectively. The average state density (in which each level with spin $J$ is counted $2J+1$ times) is then obtained by using the saddle-point approximation to the
inverse Laplace transform of the partition function~\cite{Ericson1960}, $\rho \approx \left(2\pi\beta^{-2}C\right)^{-1/2}e^{S}$.

Experiments, however, often measure the level density $\tilde{\rho}$ (in which each level is counted only once irrespective of its spin degeneracy) rather than the state density.
It was recently shown in Ref.~\onlinecite{Alhassid2013} that
level densities can be obtained directly within SMMC by calculating the density $\rho_M$ of states with given magnetic quantum number $M$ for the minimal absolute value of $M$, i.e.,
\be\label{level-density}
\tilde{\rho}  = \left \lbrace \begin{array}{lc} \rho_{M=0} & \mbox{for even-mass nuclei} \\
                                                   \rho_{M=1/2} & \mbox{for odd-mass nuclei} \end{array} \right . \; .
\ee
The density $\rho_M$ can be calculated by projection on the total spin component $M$~\cite{Alhassid2007}. Using the saddle-point approximation for the $M$-projected density, we find (in analogy with the total state density)
$\rho_M \approx \left(2\pi\beta^{-2}C_M\right)^{-1/2}e^{S_M}$,
where $S_M$ and $C_M$ are, respectively, the $M$-projected entropy and heat capacity. In this work we use this method to calculate level densities.  The $M$-projected heat capacity  $C_M$ is calculated inside the HS path integral by the method of Ref.~\onlinecite{Liu2001}, taking into account correlated errors and thus reducing significantly the statistical errors in the numerical derivative of $E_M$.

We use the CI shell model Hamiltonian described in Ref.~\onlinecite{Nakada1997} in the complete $pfg_{9/2}$ shell. This Hamiltonian is known to provide a good description of the statistical and collective properties of medium-mass nuclei in the iron region. Its interaction includes the dominating collective components of realistic nuclear  effective interactions~\cite{Dufour1996}, yet has the advantage that it satisfies the modified sign rule~\cite{Alhassid1994}, and is therefore free from the Monte Carlo sign problem for even-even nuclei.

\emph{Accurate determination of ground-state energies.}
In order to compare the SMMC level densities with experimental data, it is necessary to express them as a function of the excitation energy. This requires an accurate determination of the ground-state energy $E_0$.

For even-even medium-mass nuclei there are two accurate methods to calculate $E_0$.  The first method is based on $E_0$ being the limiting value of $E(\beta)$ for $\beta \to \infty$.  At large $\beta$, the matrix representing the propagator $\hat U_\sigma$ in the single-particle space becomes ill-defined, and we use the stabilization method of Ref.~\onlinecite{Alhassid2008} in the framework of the canonical ensemble. We can then calculate $E(\beta)$ for several large but finite values of $\beta$ and take an average.  The second method is based on the two-level model~\cite{Nakada1998}. At large $\beta$, the thermodynamic observables of the even-even nucleus are well-approximated by considering just the lowest two levels: the $0^+$ ground state and the lowest excited $2^+$ level. In this model the thermal energy $E(\beta)= \lb \hat H \rb$ and $\lb \hat{\bf  J}^2 \rb_\beta$ (where $\hat{\bf J}$ is the total nuclear spin) are (at a given value of $\beta$) functions of $E_0$ and the excitation energy $E_x^{2^+}$ of the lowest $2^+$ state.  We can then solve these relations to determine $E_0$ and $E_x^{2^+}$  from the calculated SMMC values of $E(\beta)$ and $\lb \hat{\bf J}^2 \rb_\beta$. The final estimates for $E_0$ and $E_x^{2^+}$  are then obtained by taking their average over moderate to large values of $\beta$ (for which the two-level model is valid).

\begin{figure}[t!]
    \includegraphics[width=\columnwidth]{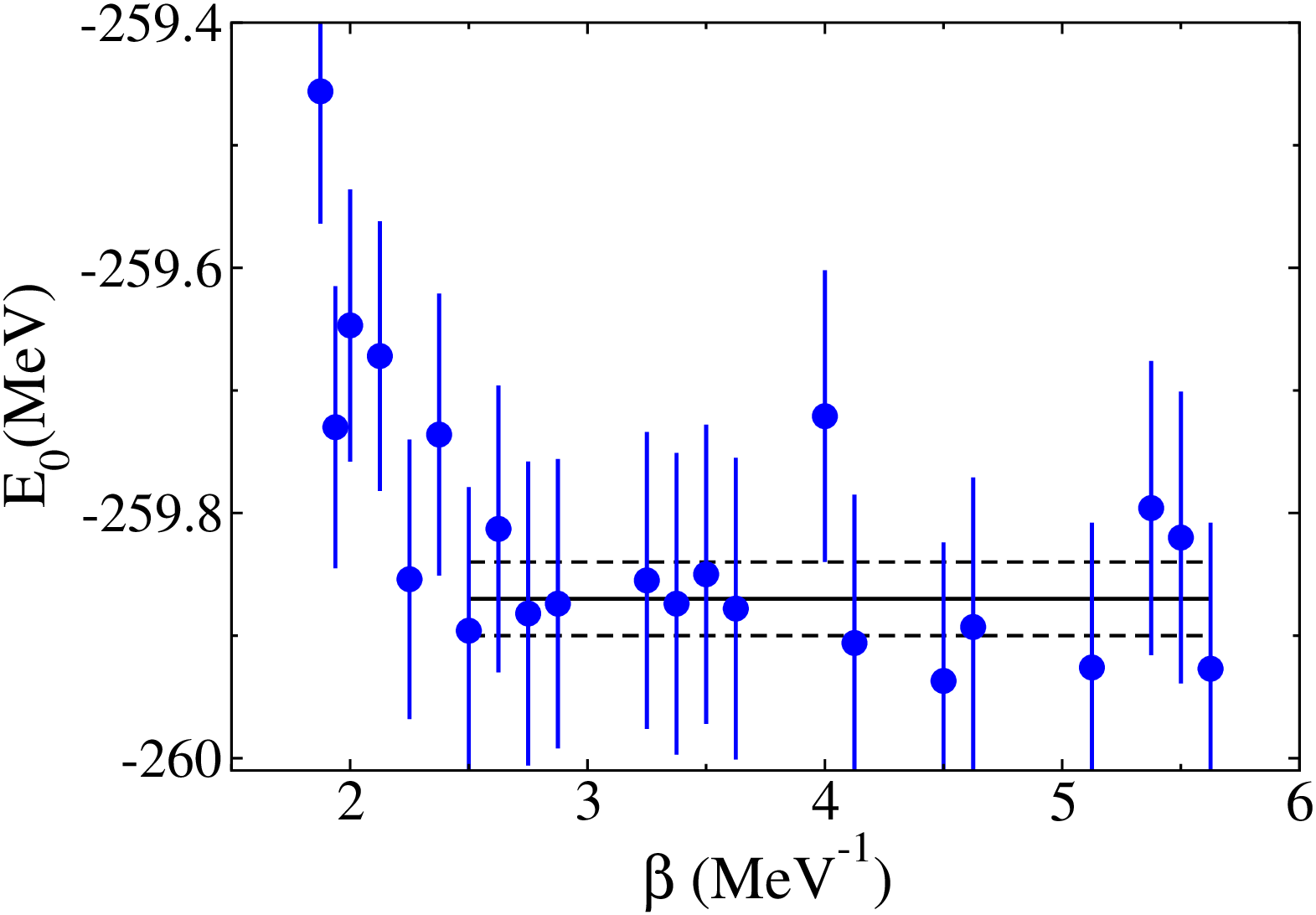}
    \caption{The parameter $E_0$ (solid circles) versus $\beta$ for $^{60}$Ni in the two-level model (see text). The solid line is our estimate for the ground-state energy obtained by averaging $E_0$ over the $\beta$ values within the length of the solid line. The dashed lines describe the statistical error in this estimate.}
    \label{Ni60-gs-2}
 \end{figure}

We demonstrate the second method in Fig.~\ref{Ni60-gs-2} for  $^{60}$Ni, where the parameter $E_0$ is shown as a function of $\beta$ (solid circles). The final estimate for the ground-state energy is the solid line obtained by averaging  $E_0$ over the range of $\beta$ values spanned by the solid line. The dashed lines describe the statistical error of this estimate.

The projection on an odd number of particles leads to a new sign problem at low temperatures, even for the good-sign interaction used in this work.  This makes it impractical to estimate the ground-state energy of an even-odd nucleus from direct SMMC calculations.
It was shown recently in Ref.~\onlinecite{Mukherjee2012} that this sign problem can be circumvented for the ground-state energy by exploiting  the asymptotic properties of the single-particle Green's functions of the neighboring even-even nuclei.

For a rotationally invariant and time-independent Hamiltonian, the scalar imaginary-time single-particle Green's functions are defined as
\be
\label{green}
G_{\nu}(\tau) = \frac{\Tr_{\ma}\left[~e^{-\beta \hat H} \mathcal{T} \sum_m a_{\nu m}(\tau) \ad_{\nu m}(0)\right] }{\Tr_{\ma}~e^{-\beta \hat H}}\;,
\ee
where $\nu \equiv (n l j)$ labels the nucleon single-particle orbital with radial quantum number $n$, orbital angular momentum $l$ and total spin $j$. Here $\mathcal{T}$ denotes
time ordering and $a_{\nu m}(\tau)\equiv e^{\tau \hat H} a_{\nu m} e^{-\tau \hat H}$ is an annihilation operator of a nucleon at imaginary time $\tau$ ($-\beta \leq \tau\leq \beta$) in a single-particle state with orbital $\nu$ and magnetic quantum number $m$ ($-j\leq m \leq j$). Using the HS transformation, the Green's functions in Eq.~(\ref{green}) can be written in a form suitable for SMMC calculations~\cite{Mukherjee2012}.

In the limit $\beta \to \infty$ and $|\tau| \to \infty$ while $|\tau| \ll \beta$, the Green's function for an even-even nucleus with a spin $J=0$
ground state is well approximated by a single exponential,
\be
\label{G-asymptotic}
G_{\nu}(\tau) \sim  e^{- \Delta E_j |\tau|}\, .
\ee
where $\Delta E_{j}$ is the energy difference between the lowest spin $J=j$ level of the relevant odd-mass nucleus and the ground state of the neighboring even-even nucleus.
In this asymptotic regime for $\tau$, we can calculate $\Delta E_j$ from the slope of $\ln G_{\nu}(\tau)$. Minimizing $\Delta E_{j}$ over all possible values of $j$, we determine the difference between the ground-state energy of the odd-mass nucleus and the ground-state energy of the even-even nucleus as well as the ground-state spin of the odd-mass nucleus.

For an even-even nucleus with a good-sign interaction, $G_{\nu}(\tau)$, and hence $ \Delta E_j $, can be calculated within SMMC free of any sign problem. Then, knowing
the ground-state energy of the even-even nucleus (which can be calculated using the method described above), the ground-state energy of the neighboring odd-mass nucleus can be calculated accurately.

In a shell model Hamiltonian with well separated energy levels, only a few transitions between the even-even and odd-mass nuclei contribute significantly to the Green's functions. In this case, the asymptotic region can be accessed even at moderate values of $\beta$.

\begin{figure}[t!]
    \includegraphics[width= \columnwidth]{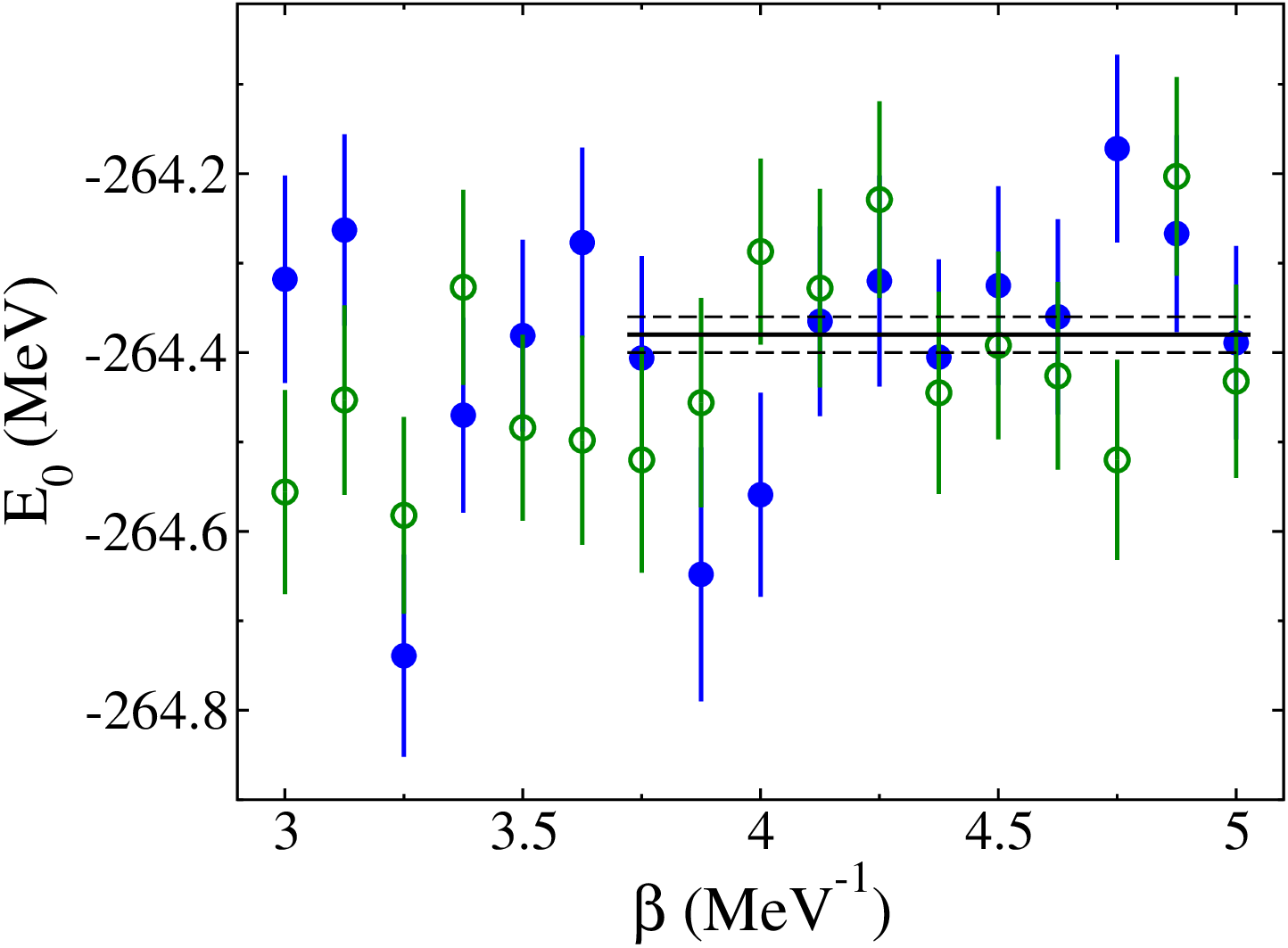}
    \caption{Ground-state energy estimate $E_0$ versus $\beta$ for $^{61}$Ni, obtained from the Green's functions of $^{60}$Ni (solid circles) and of $^{62}$Ni (open circles). Our final estimate for the ground-state energy is described by the solid and dashed lines as in Fig.~\ref{Ni60-gs-2}.}
    \label{Ni61-gs}
 \end{figure}

We demonstrate the method for $^{61}$Ni in Fig.~{\ref{Ni61-gs}}, where we show the ground-state energy $E_0$ versus $\beta$ for $^{61}$Ni. Values of $E_0$ can be determined from both the Green's functions of $^{60}$Ni (solid circles) and of $^{62}$Ni (open circles) by using Eq.~(\ref{G-asymptotic}).  The final estimate for the ground-state energy of the odd-mass nucleus is obtained by averaging over a range of $\beta$ values and is described by the solid and dashed lines (see caption of Fig.~\ref{Ni61-gs}). The statistical errors in this method are comparable to those of the even-even nuclei and are an order of magnitude smaller than the statistical errors of odd-mass nuclei in direct SMMC calculations.

Table \ref{table:E0} summarizes our results for the ground-state energies $E_0$ of both the even -mass and odd-mass nickel isotopes. Also shown for the even-mass isotopes are the extracted excitation energies of the lowest $2^+$ levels. They are in reasonable agreement with the experimental values.
\begin{table}[h!]
\begin{ruledtabular}
\begin{tabular}{lccc}
  & $E_0$ & $E_x^{2^+}$ & Exp.\\
  & (MeV) &  (MeV) &  (MeV) \\
\hline
\noalign{\smallskip}
$^{59}$Ni & -251.73(1) & --         & --  \\
$^{60}$Ni & -259.87(3) &  1.20(7)   & 1.33\\
$^{61}$Ni & -264.38(1) & --	    & --  \\
$^{62}$Ni & -271.84(3) &  1.29(8)   & 1.17\\
$^{63}$Ni & -275.67(1) & --	    & -- \\
$^{64}$Ni & -282.39(3) &  1.4(2)    & 1.35

\end{tabular}
\end{ruledtabular}
\caption{SMMC ground-state energies $E_0$  (second column) for the corresponding nickel isotopes shown in the first column. The third and fourth columns show the theoretical and experimental excitation energies $E_x^{2^+}$ of the first excited $2^+$ states in the even-mass nickel isotopes.
The theoretical values of $E_0$ and $E_x^{2^+}$ for the even-mass nickel isotopes were determined from the SMMC calculations of $E(\beta)$ and $\lb J^2\rb_\beta$ using the two-level model~\cite{Nakada1998} and taking an average over a suitable range of $\beta$ values.}
\label{table:E0}
\end{table}

\begin{figure*}[ht!]
   \includegraphics[width= 0.9\textwidth]{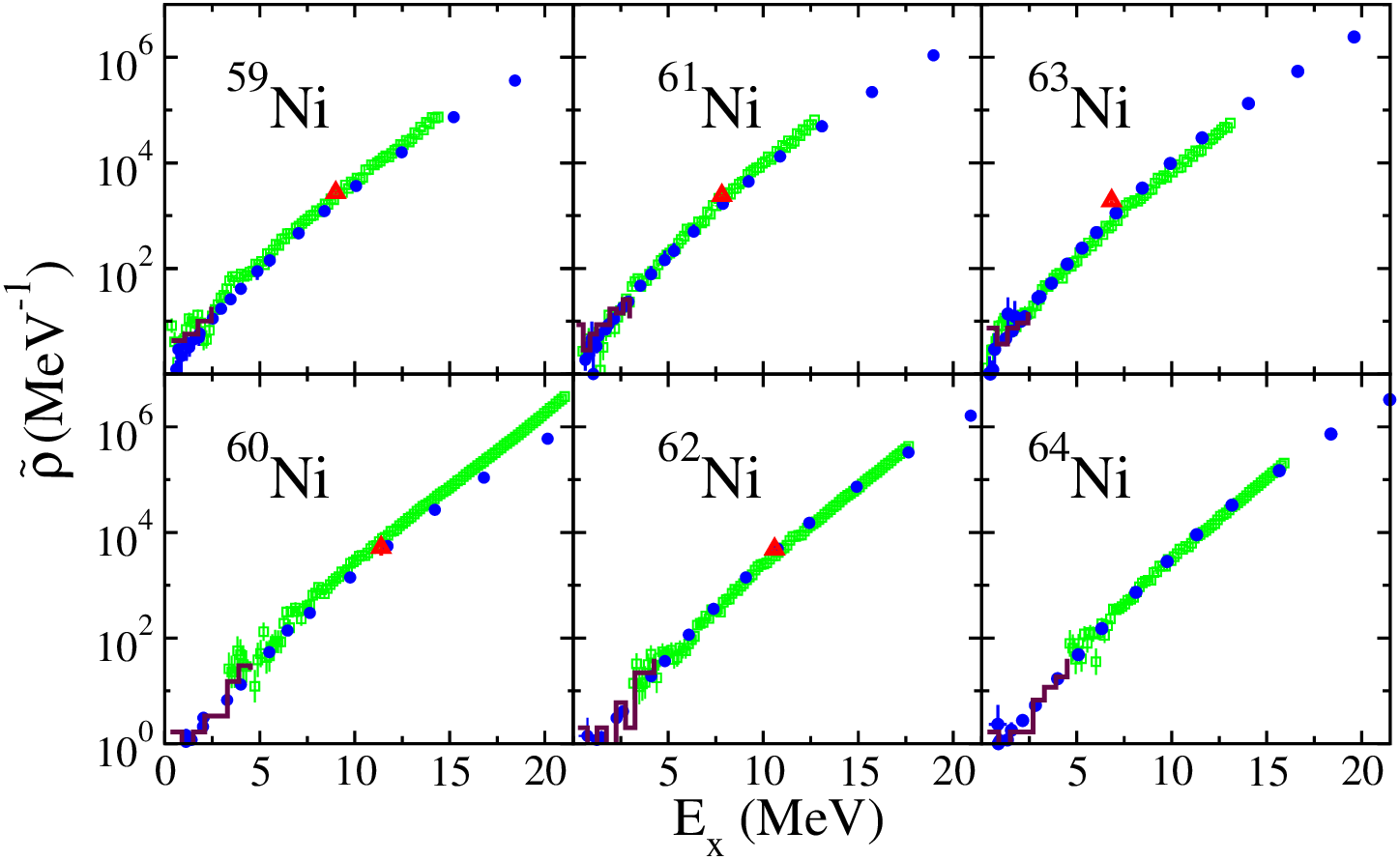}
   \caption{Level densities for $^{59-64}$Ni isotopes: theory versus experiment. The SMMC level densities (solid circles) are compared with level counting data (solid histograms)~\cite{Ripl3}, and with level densities extracted from proton evaporation spectra (open squares)~\cite{Voinov2012}, and (when available) neutron resonance data (triangle)~\cite{Ripl3}.}
 \label{Ni-lev}
 \end{figure*}

\emph{Level densities of nickel isotopes.}
The level densities for $^{59-64}$Ni are shown Fig.~\ref{Ni-lev}. The SMMC level densities (solid circles with statistical error bars) are calculated using Eq.~(\ref{level-density}).  The excitation energies are calculated from $E_x=E-E_0$ using the ground-state energies $E_0$ listed in Table \ref{table:E0}.

We compare our theoretical level densities with experimental level densities extracted from different data sets: (i)  direct level counting (solid histograms)~\cite{Ripl3}, (ii) proton evaporation spectra (open squares)~\cite{Voinov2012} and (iii) neutron resonance data (triangle)~\cite{Ripl3}.

Level counting is carried out up to a nucleus-dependent cutoff energy below which a complete set of energy levels is available from spectroscopic experiments. In Table~\ref{tbl:lvlEn} we list this cutoff energy and corresponding total number of levels used in the direct level counting for each of the nuclei. The level densities were obtained using bins of size $\sim 0.8$ MeV.
\begin{table}[ht!]
\begin{ruledtabular}
\begin{tabular}{lcccc}
	& No. of levels &  Cutoff energy &  $E_n$ &	 D\\
	& (complete set)  &	(MeV)	  	  &  (MeV)  & (eV)\\
\hline
\noalign{\smallskip}
$^{59}$Ni & 27 & 2.713 &  8.999  & 13.4(9)\\
$^{60}$Ni & 48 & 4.613 & 11.388 & 2.0(7)\\
$^{61}$Ni & 36 & 2.905 &  7.820  & 13.8(9)\\
$^{62}$Ni & 49 & 4.503 & 10.596 & 2.1(2)\\
$^{63}$Ni & 23 & 2.697 &  6.838  & 16(3)\\
$^{64}$Ni & 49 & 4.762 & -- 	      & --
\end{tabular}
\end{ruledtabular}
\caption{Total number of levels used for level counting (second column) in a complete set of experimental levels up to a certain cutoff energy (third column). The neutron resonance energy  $E_n$ and the mean $s$-wave resonance spacing~\cite{Ripl3} are shown in the fourth and fifth columns, respectively.}
\label{tbl:lvlEn}
\end{table}

Recently the Ohio University group~\cite{Voinov2012} extracted level densities of the above nickel isotopes in the intermediate energy regime from the measurements of proton evaporation spectra in $^{6,7}$Li induced reactions on $^{54,56,58}$Fe. These experimental level densities are normalized using the level counting data and are shown by open squares in Fig.~\ref{Ni-lev}.

The level density at the neutron binding energy $E_n$ is obtained from the mean spacing $D$ of $s$-wave resonances (the values of $E_n$ and $D$, where available, are shown in Table~\ref{tbl:lvlEn}). The conservation of spin and parity implies that only levels with certain spins and parities contribute to $D$. Therefore, to convert the neutron resonance data to a total level density it is necessary to make certain assumptions regarding the distributions of spin and parity. We make the usual assumptions that positive and negative parity levels contribute equally, and that the spin distribution is described by the spin-cutoff model~\cite{Ericson1960} with rigid-body moment of inertia. The corresponding level densities at the neutron binding energy are shown in Fig.~\ref{Ni-lev} by triangles.

Overall, our SMMC results are in excellent agreement with the experimental level densities over the complete experimental energy range. The SMMC level densities slightly underestimate the experimental densities in $^{59}$Ni and $^{60}$Ni. In $^{63}$Ni, our calculations are in close agreement with the level density extracted from the proton evaporation data, but are below the level density extracted from the neutron resonance data. We note, however, that the latter also differs from the proton evaporation results, as was already discussed in Ref.~\onlinecite{Voinov2012}.
The reason for this discrepancy might be that the assumption of parity equilibration, used to extract level densities from the neutron resonance data, is not entirely justified at the neutron binding energy~\cite{Nakada1997, Alhassid2000}.

\emph{Conclusion.} In conclusion, we have presented accurate microscopic calculations of the total level densities of nickel isotopes $^{59-64}$Ni using the SMMC approach. The calculations for the odd-mass nickel isotopes were made possible by using two recently developed techniques in SMMC. The ground-state energies of the odd-mass isotopes were calculated using the Green's function method of Ref.~\onlinecite{Mukherjee2012}, while the level densities were calculated directly using the spin-projection technique described in Ref.~\onlinecite{Alhassid2013}.

Our results are in close agreement with experimental level densities obtained from level counting and neutron resonance data, as well as the more recent level densities extracted from proton evaporation spectra. It will be interesting to apply the formalism developed in this work to other mass and energy regimes that are relevant in stellar nucleosynthesis and are not yet accessible in experiments.

\begin{acknowledgments}
We thank A. V. Voinov and S. M. Grimes for providing us with the experimental level densities of the nickel isotopes. This work was supported in part by the U.S. DOE grant No. DE-FG02-91ER40608. Computational cycles were provided by  the facilities of the Yale University Faculty of Arts and Sciences High Performance Computing Center.
\end{acknowledgments}

\end{document}